\documentclass[aps,pre,10pt,twocolumn,twoside,a4paper,floatfix,showkeys,showpacs]{revtex4-1}

\usepackage{graphicx}
\usepackage{amsmath,amssymb,amsfonts}
\usepackage{mathtools}
\usepackage{bm}

\usepackage[amsmath,hyperref,thmmarks]{ntheorem}

\usepackage{siunitx}
\usepackage{booktabs} 

\usepackage{time}

\newcommand{\reals}{\ensuremath{\mathbb{R}}}

\newcommand{\D}{\mathrm{d}}

\newcommand{\vertices}{\ensuremath{\mathcal{V}}}

\newcommand{\tredges}{\ensuremath{ {\mathcal{T}} }}  
\newcommand{\treespace}{\ensuremath{\reals^{\tredges}}}
\newcommand{\diredges}{\ensuremath{ {\mathcal{E}_\mathrm{d}} }}
\newcommand{\oredges}{\ensuremath{ {\mathcal{E}_\mathrm{o}} }}
\newcommand{\currents}{\ensuremath{\mathcal{O}}}
\newcommand{\tray}{\ensuremath{\gamma}}
\newcommand{\obs}{\ensuremath{\varphi}}
\newcommand{\oobs}{\ensuremath{\psi}}

\newcommand{\potentials}{\ensuremath{\mathcal{U}}}
\newcommand{\chords}{\ensuremath{\mathcal{H}}}
\newcommand{\chordspace}{\ensuremath{ {\reals^{\chords}} }}
\newcommand{\W}{\ensuremath{\mathbb{W}}}
\newcommand{\cycles}{\ensuremath{\mathcal{Z}}}
\newcommand{\cocycles}{\ensuremath{\mathcal{Y}}}
\newcommand{\del}{\ensuremath{\partial}}

\newcommand{\boundary}{\ensuremath{\del}}
\newcommand{\coboundary}{\ensuremath{\del^{*}}}
\newcommand{\abs}[1]{\ensuremath{\left|#1\right|}}
\newcommand{\assure}{\text{a.\,s.\@{}}}
\DeclareMathOperator{\Var}{Var}

\DeclareMathOperator{\img}{im}
\DeclareMathOperator{\sgn}{sgn}
\DeclareMathOperator{\Prob}{Prob}
\newcommand{\circu}[2]{\ensuremath{\mathring{#1}_{#2}}}

\newcommand{\ie}{{\textit{i.\,e.\@{}}}}
\newcommand{\eg}{{\textit{e.\,g.\@{}}}}
\newcommand{\cf}{{\textit{cf.\@{}}}}

\renewcommand{\vec}{\bm}		

\newcommand{\tmat}{\mathbb{W}}

\theoremstyle{plain}
\theoremheaderfont{\normalfont\bfseries}
\theorembodyfont{\slshape}
\theoremseparator{}
\theoremsymbol{}

\theoremstyle{nonumberplain}

\theoremstyle{plain}
\theoremheaderfont{\normalfont\bfseries}
\theorembodyfont{\rmfamily}
\theoremseparator{}

\theoremheaderfont{\scshape}
\theorembodyfont{\upshape}
\theoremstyle{nonumberplain}
\theoremseparator{}
\theoremsymbol{\ensuremath{\square}}
\newtheorem{Proof}{Proof}

\theoremstyle{plain}
\theoremheaderfont{\normalfont\bfseries}\theorembodyfont{\upshape}
\theoremseparator{}
\theoremsymbol{}

\newcommand{\revi}[1]{#1}

\newcommand\comment[1]{}

\newcommand{\eq}[1]{(\ref{eq:#1})}
\newcommand{\Eq}[1]{Eq.~\eq{#1}}
\newcommand{\Eqs}[1]{Eqs.~\eq{#1}}
\newcommand{\EQ}[1]{Equation~\eq{#1}}

\newcommand{\fig}[1]{\ref{fig:#1}}
\newcommand{\Fig}[1]{Fig.~\fig{#1}}

\newcommand{\Sec}[1]{Sec.~\ref{sec:#1}}
\newcommand{\SEC}[1]{Section~\ref{sec:#1}}

\newcommand{\App}[1]{App.~\ref{sec:#1}}

\include{revision}

\begin{document}

\date{\today{}}

\title{Fluctuating Currents in Stochastic Thermodynamics I.\\
  Gauge Invariance of Asymptotic Statistics}

\author{Artur Wachtel}
\thanks{Present address: Complex Systems and Statistical Mechanics, Physics and Materials Science Research Unit, University of Luxembourg, Luxembourg}
\author{J\"{u}rgen Vollmer}
\author{Bernhard Altaner}
\affiliation{Max Planck Institute for Dynamics and Self-Organization (MPI DS), Am Fassberg 17, 37077 G\"{o}ttingen, Germany}
\affiliation{Institute for Nonlinear Dynamics, Faculty of Physics, Georg-August University G\"{o}ttingen, 37077 G\"{o}ttingen, Germany}

\begin{abstract}
 Stochastic Thermodynamics uses Markovian jump processes to model random transitions between observable mesoscopic states.
 Physical currents are obtained from anti-symmetric jump observables defined on the edges of the graph representing the network of states.
 The asymptotic statistics of such currents are characterized by scaled cumulants.
 In the present work, we use the algebraic and topological structure of Markovian models to prove a gauge invariance of the scaled cumulant-generating function. 
 Exploiting this invariance yields an efficient algorithm for practical calculations of asymptotic averages and correlation integrals.
 We discuss how our approach generalizes the Schnakenberg decomposition of the average entropy-production rate, and how it unifies previous work.
 The application of our results to concrete models is presented in an accompanying publication.
\end{abstract}

\maketitle

\section{Introduction}
\label{sec:intro}

Stochastic Thermodynamics provides a framework for a thermodynamically consistent treatment of physical systems modeled by Markov processes \cite{Hill1966,Schnakenberg1976,Sekimoto1998,Lebowitz+Spohn1999,Seifert2005}.
\revi{Stationary} currents in these models are the hallmark of \revi{sustained} non-equilibrium conditions~\citep{Groot1984}:
they arise whenever a Markovian stochastic system does not satisfy detailed balance.
The fluctuations of these currents are interesting from both an experimental~\citep{Bustamante.etal2005,Ritort2006} and theoretical point of view (\cf{}~\citep{Seifert2012} and references therein).
Their asymptotic statistics for long times are described by Large Deviation Theory~\citep{Touchette2009,Ellis2006,Jiang2004}.
However, an analytical determination of the large-deviation rate function is -- apart from very simple or very symmetric systems, \cf{} \citep{1998PRL-Derrida.Lebowitz,Lebowitz+Spohn1999,Lau.etal2007,Verley.etal2014} -- challenging if not impossible.
In most cases, it is already difficult to find explicit expressions for the steady-state probability distribution.
This is particularly true if the transition rates of a given model depend non-linearly on multiple physical parameters.

Here, we offer an approach to systematically characterize the statistics of fluctuating currents: 
specifically, we present a method to calculate the large-deviation properties of the fluctuating currents in the form of scaled cumulants.
The set of scaled cumulants will be referred to as the \emph{fluctuation spectrum}.
For models with finite state space, we show that the fluctuation spectra of currents may be calculated without the need to solve non-linear equations.
We explicitly show how the linear structure of current-like observables carries over to the fluctuation spectra.
This allows us to express the statistics of all possible currents by the statistics of a finite set of basis elements.
Moreover, our analysis shows that this linear representation can be simplified further:
Our main result states that the asymptotic statistics of fluctuating currents are fully determined by topological cycles of the system.
Consequently, the same observable properties are produced by an entire equivalence class of observables.
This gauge invariance of the theory produces convenient expressions for the fluctuation spectra of arbitrary currents.
It generalizes the observations of Hill~\citep{Hill1977} and Schnakenberg~\citep{Schnakenberg1976} about the connection of topology and steady-state dynamics to the full counting statistics of physical currents.

The practical application of our representation is an efficient method to explore the parameter dependence of the statistics of fluctuating currents in physical models. 
We explicitly treat expectation values and time-correlation (Green--Kubo) integrals of fluctuating currents.
The potential of the method is demonstrated in an accompanying paper, where we demonstrate the effectiveness of our results for models of the molecular motor kinesin~\citep{AltanerWachtelVollmer2015}.

The present paper is structured as follows.
In \Sec{currents} we introduce fluctuating currents for Markov jump processes as our central element of study.
Large Deviation Theory is used to quantify their asymptotic statistics in the form of scaled cumulants.
\SEC{structure} concerns the algebraic structure of the scaled cumulants resulting from decompositions of the space of observables.
In \Sec{gauge} we use these decompositions to introduce and prove a gauge invariance of the fluctuation spectra.
It entails a very convenient representation of the fluctuation spectra.
Finally in \Sec{discussion} we summarize our results and relate them to other recent results on fluctuations of currents in Markov processes.
Moreover, we provide an outlook to prospective applications.
Some additional mathematical background concerning algebraic graph theory is summarized in the appendix.

\section{Asymptotic Statistics of Fluctuating Currents}
\label{sec:currents}

In the present section we revisit the formalism describing the time evolution of observables defined for ergodic Markov processes.
The statistics of their time-averaged currents can be characterized by Large Deviation Theory, 
and we will show how all scaled cumulants characterizing these distributions can explicitly be calculated by the Implicit Function Theorem.

\subsection{Currents for Markov jump processes}
\label{sec:markov}

The set 
 \(\vertices=\left\{ v_1, v_2, \dots, v_N \right\}\)
of potential outcomes of measurements on a physical system will be referred to as \emph{state space}.
Throughout this text,
 \(N\coloneqq\abs{\vertices}\)
shall be the number of possible states.
It will always be finite.
Further, we assume that at any given time the future evolution of the system only depends on its present state -- there is no memory of the past evolution -- and that the probability to encounter a certain succession of states is the same at all times. 
Hence, the evolution amounts to a time-homogeneous \emph{Markovian} jump process on \(\vertices\):
any transition from a state \(v_i\in\vertices\) to a different state
 \(v_j\in\vertices, j\neq i\)
happens stochastically with a time-independent rate \(w^i_j\geq 0\).
Moreover, we only consider Markov jump processes with \emph{dynamical reversibility} \citep{Schnakenberg1976}:
if a transition from \(v_i\) to \(v_j\) is possible so shall be the reverse transition, \ie{}
 \(w^i_j>0 \Leftrightarrow w^j_i>0\).
This condition is necessary to apply Stochastic Thermodynamics \cite{Lebowitz+Spohn1999,Maes2004,Seifert2005}.

Starting from a state \(\tray_0\in\vertices\), a \emph{realization} or \emph{random trajectory}, \(\tray\), of the Markov process is a sequence of states.
During a time \(T\) a trajectory \( \tray = \left( \tray_0, \tray_1, \dots, \tray_{n(T)} \right)\) makes a random number \(n(T)\) of jumps. 
For Markovian systems  an ensemble is characterized by an initial probability distribution
 \(\vec{p}(0)=\left( p_1(0), p_2(0), \dots, p_N(0) \right)\)
on \(\vertices\).
It evolves according to the master equation~\citep{vKampen1992}
\begin{align}
 \frac{\D}{\D t} p_j(t) = \sum_{\substack{i=1\\
   i\neq j}}^{N}\left(p_i(t) w^i_j - p_j(t) w^j_i\right) = \sum_{i=1}^{N} p_i(t)\, w^i_j\,, \label{eq:master}
\end{align}
where
 \(w^j_j\coloneqq-\sum_{i:\,i\neq j} w^j_i\).
Note that the master equation guarantees conservation of probability, \ie{}
 \(\sum_j p_j(t) \equiv 1\)
for all \(t\geq 0\).

The numbers \(w^i_j\) defined in \Eq{master} can be gathered in a square matrix, \(\W\), yielding the master equation in matrix form as
 \(\dot{\vec{p}} = \vec{p} \W\).
Henceforth, we assume \(\W\) to be irreducible, meaning that any two states are connected by a finite trajectory.
Together with the assumption of dynamical reversibility, this ensures the existence of a unique left-eigenvector \(\vec\pi\) of \(\W\) satisfying \(\vec 0=\vec\pi \W\) and \(\sum_i \pi_i = 1\).
This vector \(\vec{\pi}\) is called \emph{steady-state distribution} or \emph{ergodic measure} of the Markov process: all initial probability distributions, \(\vec p(0)\), relax to \(\vec{\pi}\) \citep{Feller1978}.

The \emph{steady-state probability current} for the transition from \(v_i\) to \(v_j\) is given by
 \( J^i_j \coloneqq \pi_i w^i_j - \pi_j w^j_i = - J^j_i \)
where
 \(\forall i\colon J^i_i \coloneqq 0\).
Using this current, the condition on the steady-state distribution \(\vec{\pi}\) amounts to
\[\forall j\colon 0 = \frac{\D}{\D t} \pi_j = \sum_i \pi_i w^i_j = \sum_i J^i_j\,.\]

Henceforth, we are interested in the jump observables \(\obs\) defined for a given Markov jump process.
A jump observable associates a value \(\obs^i_j\in\reals\) to a transition from one state \(v_i\) to another state \(v_j\).
Further, a jump observable needs to be balanced:
summing the values of \(\obs\) along a realization \(\gamma\) of the Markov process should only depend on the \emph{net} number of transitions between two states, requiring that \(\obs^i_j = - \obs^j_i\).
For any non-admissible transition with a vanishing transition rate  \(w^i_j=w^j_i = 0\) we define \( \obs^i_j =\obs^j_i\coloneqq 0\).
More abstractly, one can think of jump observables as a special kind of antisymmetric matrices  \(\obs\in\reals^N\).
A prominent example of a jump observable is the logarithmic ratio of the transition rates $\sigma^i_j=\ln(w^i_j/w^j_i)$, which expresses the thermodynamic dissipation in stochastic thermodynamics \cite{Lebowitz+Spohn1999,Seifert2005}.
Other jump observables may represent changes in particle numbers in chemical reactions, configurational changes, or differences in (thermodynamic) potentials \cite{Seifert2008}.

The \emph{time average} of a jump observable \(\obs\) along a given trajectory \(\tray=(\tray_0,\tray_1,\dots, \tray_{n(T)}) \) is defined as  
\begin{align}
  \overline{\obs}_T \coloneqq \frac{1}{T}  \sum_{k=1}^{n(T)}\obs^{\tray_{k-1}}_{\tray_k}\,.\label{eq:time-average}
\end{align}
If the trajectory \(\gamma\) stays unspecified, \Eq{time-average} defines a \emph{time-averaged fluctuating current}.
It constitutes a real-valued random variable, and
the ergodic theorem states that it
will almost surely (\assure) converge to ensemble averages~\citep{Feller1978,Jiang2004},
\begin{align}
  \lim_{T\to \infty} \overline{\obs}_T =  \sum_{i=1}^{N} \sum_{j=i}^{N} J^i_j\, \obs^i_j \quad\assure \label{eq:ergodic}
\end{align}
In the following we characterize the distribution of these time-averaged currents by calculating all scaled cumulants of the distribution of \(\overline{\obs}_T\).
Formally, they characterize the \(T \to \infty\) asymptotic behavior of the distributions.
Typically, the approach towards the asymptotic scaling form is exponentially fast, such that the asymptotic statistics characterize the fluctuations already for (sufficiently large) finite times \(T\)~\citep{Ellis2006}.

\subsection{Large Deviation Theory: quantifying asymptotic fluctuations}
\label{sec:fluctuations}

\EQ{ergodic} states that in the infinite-time limit the probability density of \(\overline{\obs}_T\) converges to a Dirac delta distribution.
In popular terms we say that fluctuations do not matter in the asymptotic limit.
For finite times $T$ its fluctuations can be quantified by means of the moments, or equivalently, the cumulants of the distribution of \(\overline{\obs}_T\).
To fix our notation, we briefly revisit the notion of cumulants before we use Large Deviation Theory~\citep{Touchette2009,Ellis2006} to quantify their scaling.

Let  \(X\) be a real-valued random variable, 
and let \(\rho_X(x)=\Prob(X=x)\) be its probability distribution.
The \emph{expectation} of a function
 \(\psi\colon \reals\to\reals\)
is
 \(\langle \psi(X) \rangle \coloneqq \int_\reals \psi(x) \rho_X(x) \D x\).
The expectation of the identity, \(\psi(x)=x\), is referred to as \emph{expectation value}, \(\langle X \rangle\), of \(X\).
The fluctuations of a random variable around its expectation value are commonly quantified by the variance,
 \(\Var X = \langle (X - \langle X \rangle)^2\rangle\).
 When the distribution is centered around the expectation value, the two numbers \(\langle X \rangle\) and \(\Var X\) provide a good estimate on the distribution.
Higher order cumulants further characterize its shape.
They take non-negligible values only when the distribution is not centered or strongly asymmetric, and they characterize correlations in the multi-variate case.
Formally, the cumulants are defined as follows:

Consider an \(\reals^d\)-valued random variable \(\vec{X}\).
Its \emph{cumulant-generating function}
 \(g_{\vec{X}}\colon \reals^d\to\reals\)
is defined as
\begin{align}
  g_{\vec{X}}(\vec{q}) \coloneqq \ln \left\langle \exp\left( \vec{q}\cdot\vec{X} \right)\right\rangle\,. \label{eq:cgf}
\end{align}
The partial derivatives of \(g_{\vec{X}}\),
\begin{align}
 \kappa(X^{(i_1)},X^{(i_2)}, \dots,X^{(i_\nu)}) \coloneqq \frac{\del}{\del q_{i_1}}\frac{\del}{\del q_{i_2}}\dots\frac{\del}{\del q_{i_\nu}} g_{\vec{X}}(\vec{0})\,,
\end{align}
evaluated in the origin are the \emph{joint cumulants}~\citep{vKampen1992}.
They are multi-linear in their arguments, even if the
 \(X^{(1)}, \dots, X^{(d)}\)
are \emph{not} independent~\citep{McCullagh1987}.
For independent random variables, the mixed joint cumulants vanish and the cumulants become purely additive.
The cumulant-generating function \(g_{\vec{X}}(\vec{q})\)
is (non-strictly) convex and satisfies
 \(g_{\vec{X}}(\vec{0})=0\)
for all \(\vec{X}\).
In the uni-variate case, \(d=1\), one defines 
\begin{align}
  \kappa(\underbrace{X, \dots, X}_{\nu\text{ times}}) \eqqcolon \kappa_\nu(X)\,.
\end{align}
Moreover, we have
\begin{align*}
  \kappa_1(X) &= \langle X \rangle \, , \quad 
  \\[1mm]
  \kappa_2(X) &= \left\langle (X-\langle X \rangle)^2\right\rangle = \Var X \, ,
  \\[1mm]
  \kappa_3(X) &= \left\langle (X - \langle X \rangle )^3\right\rangle \, .
\end{align*}
For \(\nu \geq 4\) the cumulants  \(\kappa_\nu (X)\) differ from the centered moments 
\(\left\langle (X - \langle X \rangle )^\nu\right\rangle\).
However,
there is a bijection from cumulants to (centered) moments \citep{vKampen1992,McCullagh1987}.

The time averages defined in \Eq{time-average} are families of random variables: 
for every given time \(T\), the time average \( \overline{\obs}_{T} \) is a random variable and has an associated probability density \(\rho_{\overline{\obs}_{T}}\).
In order to characterize the correlations of \(d\) different time-averaged currents, \( \overline{\obs}_{T}^{(1)}, \overline{\obs}_{T}^{(2)}, \dots, \overline{\obs}_{T}^{(d)} \), it is convenient to introduce the
vector
\(\overline{\vec{\obs}}_T=\left( \overline{\obs}_{T}^{(1)}, \overline{\obs}_{T}^{(2)}, \dots, \overline{\obs}_{T}^{(d)} \right)\).
It is a particular type of an \(\reals^d\)-valued random variable \(\vec{X}\).
In view of \Eq{ergodic} the associated density \(\rho_{\overline{\vec{\obs}}_{T}}\) converges to a $d$-dimensional Dirac delta function. Consequently, the cumulants of order greater than one decay to zero for large \(T\).

Large Deviation Theory is concerned with the rate of this decay~\citep{Touchette2009,Ellis2006}.
It asserts that the \emph{scaled cumulant-generating function} 
\begin{align}
 \lambda_{\vec{\obs}}(\vec{q}) \coloneqq \lim_{T\to\infty}\frac{1}{T}\ln \left\langle \exp[T\,\vec{q} \cdot \overline{\vec{\obs}}]\right\rangle\,
 \label{eq:def-scgf}
\end{align}
exists, where again \(\vec{q}\in\reals^d\).
This generating function is related to the (non-scaled) cumulant-generating function, \Eq{cgf}, by the scaling relation 
 \( \lambda_{\vec{\obs}}(\vec{q}) \coloneqq \lim_{T\to\infty} \frac{1}{T}\, g_{\overline{\vec{\obs}}_T}(T\vec{q}) \).
As a consequence, the scaled cumulant-generating function inherits the properties of the (non-scaled) cumulant-generating functions: convexity and
 \(\lambda_{\vec{\obs}}(\vec{0})=0\)
for every fluctuating current.

The partial derivatives of the scaled cumulant-generating function with respect to the components \(q_i\) of \(\vec q\) are the \emph{scaled cumulants}:
\begin{align}
	c(\obs^{(i_1)},\obs^{(i_2)}, \dots,\obs^{(i_\nu)}) \coloneqq \frac{\del}{\del q_{i_1}}\frac{\del}{\del q_{i_2}}\dots\frac{\del}{\del q_{i_\nu}} \lambda(\vec{0}) \,. \label{eq:def-scaled-cumulants}
\end{align}
From the definition, \Eq{def-scgf}, the scaled cumulants, \(c\), inherit multi-linearity from the cumulants, \(\kappa\).
Moreover, we immediately obtain their scaling
\begin{align}
 c(\obs^{(i_1)},\dots,\obs^{(i_\nu)}) = \lim_{T\to\infty} T^{\nu-1} \kappa( \obs^{(i_1)}_T, \dots,\obs^{(i_\nu)}_T)\,.
 \label{eq:cumulant-scaling}
\end{align}

In the following, the term \emph{fluctuation spectrum} shall refer to the set of all scaled cumulants of a given vector $\overline{\vec{\obs}}_T$ of time-averaged currents.
For time-averages \(\overline{\obs}_T\) of a single jump observable $\obs$, the fluctuation spectrum is the sequence of scaled cumulants:
\(c_\nu(\obs)=\lim_{T\to\infty}T^{\nu-1}\kappa_\nu(\overline{\obs}_T)=\lim_{T\to\infty}\frac{1}{T}\kappa_\nu(T\overline{\obs}_T)\).
Note that in the physics literature dealing with large deviations, the word ``scaled'' is often implied when speaking of ``cumulants''.
Furthermore, the joint scaled cumulants of order two are identical to asymptotic time-correlation (Green--Kubo) integrals~\citep{Lebowitz+Spohn1999}.

\subsection{Calculating the fluctuation spectrum}
\label{sec:analytics}

For ergodic Markovian jump processes with finite state space the scaled cumulant-generating function \(\lambda_{\vec{\obs}}(\vec{q})\) has an interesting property:
it is the unique dominant eigenvalue of the matrix \(\W_{\vec{\obs}}(\vec{q})\)
with components
\begin{align}
 \left(\W_{\vec{\obs}}(\vec{q})\right)^i_j \coloneqq w^i_j \exp\left( \vec{q} \cdot\vec{\obs}^i_j \right),
 \label{eq:skewed-generator}
\end{align}
where \(\vec{q}\in\reals^d\), and \(w^i_j\) is the rate of jumps from \(v_i\) to \(v_j\)~\citep{Ellis2006,Touchette2009}.

In general, it is impossible for \(N =\abs{\vertices} > 4\) 
to find analytic expressions for the dominant eigenvalue
 \(\lambda_{\vec{\obs}}(\vec q)\)
of the matrix
 \(\W_{\vec{\obs}}(\vec{q})\).
After all, the solution of the eigenvalue problem amounts to finding the root with the largest real part of a polynomial of degree \(N\).
An analytical expression for \(\lambda_{\vec{\obs}}(\vec q)\) is 
needed, however, to evaluate the (higher-order) scaled cumulants as derivatives, \Eq{def-scaled-cumulants}, of the eigenvalue.
In most cases, this is where the derivation of exact analytic results stops -- unless there are very special properties of the system, \eg{} symmetries, that help finding the eigenvalue (see \citep{1998PRL-Derrida.Lebowitz,Lau.etal2007,Verley.etal2014} for noticeable examples).

An important message of the present paper is that determining any cumulant of finite order does not require finding the solution of any non-linear equation.
All relevant information is contained in the characteristic equation
\begin{align}
  0 = \det\left(\W_{\vec{\obs}}(\vec{q}) - \lambda \mathbb{E}\right) \eqqcolon \chi_{\vec{\obs}}(\vec{q},\lambda) \eqqcolon\sum_{k=0}^{N} a_{k}(\vec{q})\,\lambda^{k}
\label{eq:characteristic}
\end{align}
solved by the eigenvalues \(\lambda\) of the matrix
 \(\W_{\vec{\obs}}(\vec{q})\).
 \revi{The above equation, together with the property of the dominant eigenvalue
 \(\lambda_{\vec{\obs}}(\vec{0})=0\),
 implicitly defines the scaled cumulant-generating function \(\lambda_{\vec{\obs}}(\vec{q})\).
 Since the characteristic polynomial is differentiable in both arguments, the Implicit Function Theorem can be applied to \Eq{characteristic} to extract the partial derivatives of \(\lambda_{\vec{\obs}}(\vec{q})\).
 The Implicit Function theorem further guarantees that the implicitly defined function is (locally) unique.
 Consequently, the thus calculated partial derivatives agree with the scaled cumulants. 
 In practice, this means that} \Eq{characteristic} provides the scaled cumulants iteratively and directly from the coefficients \(a_k(\vec{q})\) of the characteristic polynomial: take partial derivatives with respect to various \(q_\ell\) of the equation
 \( 0 = \chi_{\vec{\obs}}(\vec{q},\lambda_{\vec{\obs}}(\vec{q}))\)\,,
 and keep track of the dependence of
 \(\lambda_{\vec{\obs}}(\vec{q})\) and its derivatives on the variable \(\vec{q}\).
Evaluating these expressions at \(\vec{q}=\vec{0}\) and
 \(\lambda_{\vec{\obs}}(\vec{0})=0\)
makes the scaled cumulants appear as inner derivatives.
More importantly, they appear linearly in the expressions and thus can be solved for. 
Hence, their dependence on the (possibly parameter-dependent) transition rates $w^i_j$ and jump observables $\obs^i_j$  is analytically accessible -- in contrast to that of the dominant eigenvalue \(\lambda_{\vec{\obs}}(\vec{q})\).

Regarding the application of the Implicit Function Theorem, one has to ensure that the coefficient
\(a_1(\vec{0})=\del_{\lambda}\chi_{\vec{\obs}}(\vec{0},0)\)
does not vanish.
This is in fact given: since
 \(\W_{\vec{\obs}}(\vec{0}) = \W\),
the values \(a_k(\vec{0})=a_k\) are independent of the observable \(\vec{\obs}\).
They only depend on the transition matrix \(\W\), and the Matrix Tree Theorem~\citep{Tutte2001} ensures that
 \(a_1(\vec{0})=a_1\neq 0\)
for every irreducible transition matrix.

To be explicit, we give the general expressions for the first two scaled cumulants 
\(c\left( \obs^{(\ell)} \right)\) and \( c\left( \obs^{(\ell)},\obs^{(m)} \right)\) 
of the vector \( \overline{\vec{\obs}}_T\) 
in terms of the coefficients \( a_{k}\left( \vec{q} \right)\) of the characteristic polynomial:
\begin{widetext}
\begin{subequations}
  \begin{align}
    c(\obs^{(\ell)}) 
    &= -\frac{\partial_\ell a_0}{a_1} \label{eq:c1general} \eqqcolon c^{(\ell)}
\\[2mm]
    c\left(\obs^{(\ell)},\obs^{(m)}\right) 
    &=  -\frac{\partial^2_{\ell m} a_0}{a_1} + \frac{(\partial_m a_1)(\partial_\ell a_0)+(\partial_\ell a_1)(\partial_m a_0)}{a_1^2}\nonumber 
     -\frac{2(\partial_\ell a_0)(\partial_m a_0)a_2}{a_1^3}
    \nonumber\\[2mm]
     = &-\frac{\partial^2_{\ell m}a_0 + (\partial_\ell a_1) c^{(m)} +(\partial_m a_1)c^{(\ell)} +2 a_2 c^{(m)} c^{(\ell)}}{a_1}\label{eq:c2general}
  \end{align}
\label{eq:c1c2general}
\end{subequations}
\end{widetext}
where the partial derivatives \(\partial_\ell a_k \equiv \left.\frac{\partial a_k(\vec q)}{\partial q_\ell}\right\vert_{\vec q = \vec{0}}\) and the coefficients \(a_k\) are evaluated at \(\vec q = \vec0\).
The higher cumulants have a similar but more involved representation in terms of the coefficients \(a_k(\vec{q})\).
Note the following facts:
The coefficients up to order \(k\) are sufficient to uniquely determine the cumulants up to order \(k\).
Consequently, if one is interested only in scaled cumulants of order \(2\) and \(1\), one only needs the coefficients \(a_0(\vec{q})\equiv\det \W_{\vec{\obs}}(\vec{q})\), \(a_1(\vec{q})\) and \(a_2(\vec{q})\) -- irrespective of the degree, \(|\vertices|\equiv N\), of the characteristic polynomial.
Furthermore, the \Eqs{c1general} and \eq{c2general} are agnostic to the dimension \(d\) of the vector \(\vec{\obs}^i_j\) of jump observables.

\subsection{Summary}
\label{sec:statistics-summary}

We conclude that the Implicit Function Theorem allows us to calculate the fluctuation spectrum based on only the transition matrix \(\W\) and the observable \(\vec{\obs}\).
At no step do we make use of analytical expressions for the scaled cumulant-generating function 
 \(\lambda_{\vec{\obs}}(\vec{q})\)
or for the steady-state distribution \(\vec{\pi}\).
The combinatorics sometimes used to calculate \(\vec{\pi}\) is dealt with implicitly, and much more elegantly, by the characteristic polynomial (\cf{} Ref.~\citep{Boon2012} for explicit considerations of the combinatorics involved in a scaled cumulant of second order).
In other words, the relevant information about the steady-state distribution is hidden in the coefficients \(a_k(\vec{q})\).

In the following section, we present a method calculate the scaled cumulants in a systematic and efficient way:
due to the vector-space structure of the current observables and the multi-linearity of scaled cumulants, it is enough to calculate the cumulants of a reduced set of fundamental observables, acting as a basis.

\section{Algebraic Structure of Fluctuation Spectra}
\label{sec:structure}

\begin{figure}
 \hspace*{\fill} \raisebox{26mm}{(a)} \includegraphics[scale=1.2]{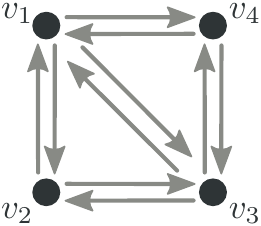} \hspace*{\fill} \raisebox{26mm}{(b)} \includegraphics[scale=1.2]{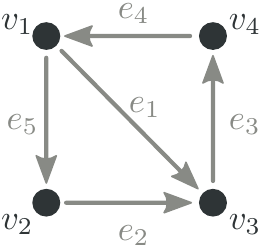} \hspace*{\fill}
 \caption{An example graph representing the network of states of a dynamically reversible Markov process, as used in models of Stochastic Thermodynamics.
	 It is represented either as (a) a graph \( \left( \vertices, \diredges \right)\) with two directed edges connecting each pair of, states or (b) by choosing a reference orientation as a simple graph \( \left( \vertices, \oredges \right)\).
 We will use the latter throughout this text.}
 \label{fig:graph}
\end{figure}

Kirchhoff established two laws characterizing the stationary distributions of electrical currents in electrical networks \citep{Kirchhoff1847}.
The current law states that steady-state currents balance at each vertex.
The voltage law states that the sum of potential differences vanishes when integrated along a cycle.
The field theory of electrodynamics reformulates these observations in the context of vector calculus. 
We will now explore a corresponding structure for Markovian jump processes.
In \Sec{gauge} we thus establish a correspondig gauge invariance for the asymptotic statistics of fluctuating currents.
Adopting the language of graph theory~\citep{Gross1987,Tutte2001,Knauer2011} allows us to do this in a concise and elegant way.

\subsection{The network of states as a graph}
\label{sec:markov-graphs}

The structure of a Markovian jump process can be thought of as a graph where the states \(\vertices\) are the \emph{vertices} or \emph{nodes}.
For every non-vanishing transition rate \(w^i_j>0\) we draw a \emph{directed edge} \((v_i\to v_j)\) that we denote by \(e^i_j\).
The set of all directed edges will be denoted as \(\diredges\) and the pair
 \( \left( \vertices,\diredges \right)\)
 is a \emph{graph}, \cf{}~\Fig{graph}(a).
The Markovian jump process performs steps along edges \(e^i_j\) of the graph with stochastic rates \(w^i_j\).

In a dynamically reversible Markov process, we have a pair of edges \( \{e^i_j,e^j_i\} \) connecting each pair of vertices \( \{v_i,v_j\} \), if transitions between these vertices are admissible.
In that case we define
 \( -e^i_j\coloneqq e^j_i\)
since the edges only differ by their direction.
For every pair of connected vertices \(\{v_i,v_j\}\) we can choose one of the edges as a reference.
We shall call this an \emph{oriented edge}.
The set \(\oredges\) of oriented edges satisfies
\(\oredges\cup -\oredges=\diredges\), see \Fig{graph}.
The Markov process defines a random walk on the graph
 \( (\vertices,\oredges) \)
where steps are also allowed 
in the reversed direction of an oriented edge \(e\in\oredges\).
Note that, in contrast to
 \( (\vertices,\diredges)\),
the graph
 \( (\vertices,\oredges) \)
is \emph{simple}: There is at most one edge connecting two vertices. 
In the following let
 \(M\coloneqq\abs{\oredges} = \frac{1}{2}\abs{\diredges}\)
be the number of connected pairs of states.
Henceforth, we enumerate the oriented edges by a single index, \ie{}
 \(\oredges=\left\{ e_1, \dots, e_M \right\}\), \cf{}~\Fig{graph}(b).

Recall that the realization of a Markovian jump process is a random walk \(\gamma\).
Instead of writing $\gamma = (\tray_0,\tray_1,\dots\tray_n)$ as a tuple of subsequent vertices, we can alternatively define a walk by its \(n\) edges and write
\(\gamma = \left( e_{i_1}, \dots, e_{i_n} \right)\).
An example of a walk from \(v_4\) to \(v_3\) on the oriented graph in \Fig{graph}(b) is the sequence
 \( \left( e_4, e_1, -e_2, -e_5, -e_4, -e_3 \right)\).

\subsection{Vector-space algebra for graphs and fluctuation spectra}
\label{sec:algebra}

Graph theory also provides a complementary perspective on the jump observables defined in \Sec{markov}.
Above, we introduced jump observables as anti-symmetric matrices with vanishing entries whenever a transition is not admissible.
In the present context, it is more appropriate to use an (equivalent) definition based on the oriented edges $\oredges$, rather than on (pairs of) vertices.
To that end, we consider the set
\begin{align}
 \currents \coloneqq\left\{ \obs\colon\oredges \to \reals\right\} \equiv \reals^\oredges
 \label{eq:jump}
\end{align}
of all real functions $\obs$ defined on the oriented edges.
As a function space it inherits a vector-space structure.
Via anti-symmetry, it naturally extends to all the directed edges \(e\in\diredges\): \( \obs(-e) \equiv -\obs(e) \)\,.
Such a structure is called (anti-symmetric) \emph{edge space} in the mathematics literature~\citep{Knauer2011}.
Henceforth, a jump observable is an element $\obs\in\currents$ of the edge space.
Note that with this definition, all the results obtained in the present paper also hold for physical models with multiple (bi-directional) edges between states, like the ones considered (for instance) in Refs.~\cite{Lau.etal2007,Esposito.VandenBroeck2010}.

There is a canonical way of defining a basis for the jump observables:
For every edge \(e\in\oredges\) the corresponding indicator function is an observable that takes the value \(1\) on this edge and \(0\) everywhere else.
For convenience, we denote this observable also by \(e\).
In this sense, the edges are observables themselves,
 \(\oredges \subset \currents\).
Every jump observable \(\obs\in\currents\) can be written as a linear combination
 \(\obs=\sum_{m=1}^{M} \obs_{m}e_m\)
 with \(\obs_{m}=\obs(e_m)\in\reals\), and thus the oriented edges serve as a basis.
The natural scalar product on \(\currents\),
\begin{align}
 \left\langle \obs, \oobs \right\rangle 
 \coloneqq \sum_{e\in\oredges} \obs(e) \oobs(e) 
 = \sum_{m=1}^{M}\obs_{m}\oobs_{m}\,,\label{eq:scalar-product}
\end{align}
treats the oriented edges \(\oredges\) as an orthonormal basis.
After all, irrespective of the choice of orientation one has
 \( \left\langle e_m, e_k\right\rangle = e_m(e_k) = \delta_{mk}\).
Consequently, the initially arbitrary orientation of the edges amounts to the choice of a basis.
Moreover, the ergodic theorem, \Eq{ergodic}, may thus be stated as 
\( \lim_{T\to\infty} \overline{\obs}_T = \langle \obs, J \rangle\)\,.

The multi-linearity of the scaled cumulants implies that the statistics of every jump observable can be described by the (joint) scaled cumulants of a basis:
 \begin{align}
  c_\nu(\obs) 
  = \left( \prod_{m=1}^\nu \sum_{e_m\in\oredges} \obs_m \right) c(e_1, e_2, \dots,e_\nu) \,.
 \label{eq:naive-multi-linear-cumulants}
 \end{align}
 Here the \(e_m\) need not be the canonical basis introduced above, but can be \emph{any} basis -- given that the \(\obs_m\) represent the coefficients of \(\obs\) in that basis.
A representation of this form is not necessarily useful for practical calculations.
The effort to calculate the fluctuation spectra for the entire basis exceeds the one of calculating the fluctuation spectrum of a single observable. 
Regardless, \Eq{naive-multi-linear-cumulants} provides valuable conceptual insights. 
On the one hand, it establishes a connection between the fluctuations of different observables, which is interesting in its own right.
On the other hand, we will see later that a practical computational algorithm emerges through an informed choice of the basis.
In the remainder of the present section we further explore the linear structure of the vector space \(\currents\).
In \Sec{gauge} we will combine these insights to establish a gauge invariance for the fluctuation spectra that will allow us to choose a basis minimizing the effort involved in calculating scaled cumulants.

\subsection{Cycles and co-cycles}
\label{sec:cycles-cocycles}

In order to formulate the analogy between Markov processes and electrodynamics, we consider jump observables as ``discretized vector fields'' that point along the edges of the graph. 
The jump observables are the elements of the edge space \(\currents=\reals^{\oredges}\), \ie{} (anti-symmetric) real functions on the edges.
The analogous notion of ``scalar fields on discrete sets'' is the so-called \emph{vertex space}, \ie{} the space of real functions on the vertices
 \(\potentials\coloneqq\reals^{\vertices}=\left\{ u\colon \vertices\to\reals \right\}\).
We identify the vertex set \(\vertices\) with its indicator functions and regard the vertex space \(\potentials\) as the linear span of \(\vertices\).
This makes the vertices a subset
 \(\vertices\subset\potentials\)
of the vertex space, just as we regard
 \(\oredges\subset\currents\).
Again, the natural scalar product on \(\potentials\) treats the basis \(\vertices\) as orthonormal.

\begin{figure}
 \centering \includegraphics[width=.48\textwidth]{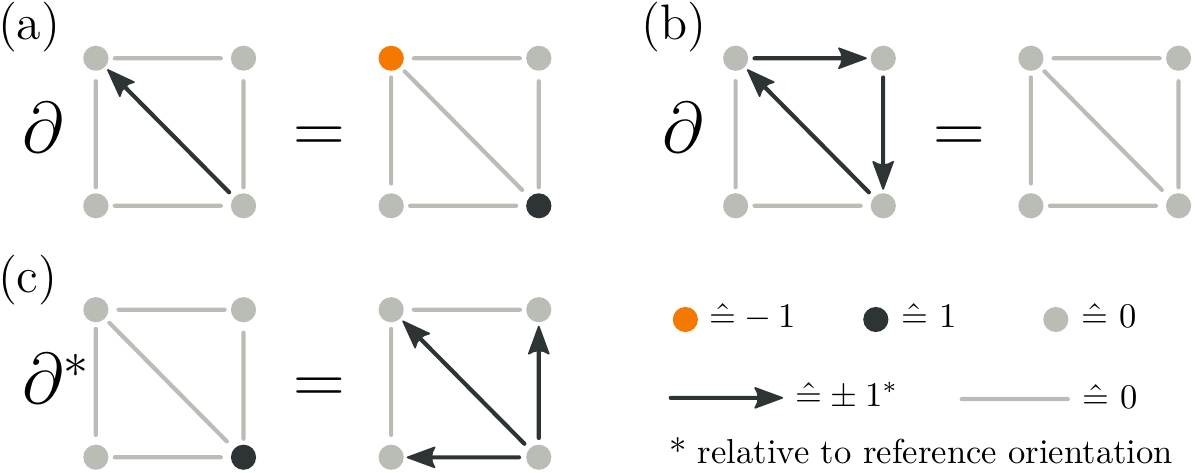}
 \caption{\revi{The action of the boundary and the co-boundary exemplified using the graph of \Fig{graph}(b). 
 The reference orientation is omitted here.
 (a) The boundary operator \(\boundary\) maps an edge (left) to a linear combination of the two vertices that are connected by the edge (right).
 (b) A linear combination of edges that does not have any sinks or sources has boundary zero -- the defining property of a \emph{cycle}, see main text.
 (c) The co-boundary operator \(\coboundary\) maps a single vertex (left) to the sum of edges pointing away from that vertex (right). 
  The coefficients for the linear combinations of edges and vertices are encoded as indicated at the bottom right.}
  Note that a negative coefficient associated to an edge is equivalent to a positive coefficient with reversed orientation.
 }
 \label{fig:boundary-coboundary}
\end{figure}

Analogously to vector calculus, there are natural difference operators that map vector fields to scalar fields, and vice versa:
The \emph{boundary operator} \(\boundary\colon \currents\to\potentials\) and the \emph{co-boundary operator} \(\coboundary\colon \potentials\to\currents\), respectively.
On the basis elements they act as \(\boundary \colon e^i_j \mapsto v_i-v_j\) and \(\coboundary\colon v_i \mapsto \sum_{j: j\neq i} e^{i}_{j}\)\,, \revi{ \cf{} \Fig{boundary-coboundary}(a) and (c), respectively.
Being linear operators they commute with sums, \ie{} with any linear combinations of edges and vertices.}
It is easy to check that, as notation suggests, they are dual operators with respect to the natural scalar products.
\revi{As exemplified in \Fig{boundary-coboundary}, these linear difference operators have a very natural interpretation when jump observables are understood as discrete vector fields:} the operator \(\boundary\) acts like a discrete divergence, while \(\coboundary\) resembles a negative discrete gradient.

Kirchhoff's current law states that the net current into each vertex of an electrical network balances the net outflow.
In abstract terms, the formulation of this statement applied to the probability current \(J\) reads \(\boundary J=0\).
It is equivalent to the steady-state master equation~\eqref{eq:master}, 
 \(\vec{0}=\vec{\pi}\mathbb{W}\).
Another way to formulate the result is that probability (or electrical charge) is conserved and there are neither sinks nor sources.
Then, such currents must run in cycles and we define the \emph{cycle space} as
\(\cycles\coloneqq \ker \boundary\), \revi{\cf{} \Fig{boundary-coboundary}(b) for an example of a cycle}.
In the picture of discrete vector fields, these are ``divergence-free fields''.
The ``gradient fields'' form the \emph{co-cycle space}
 \(\cocycles\coloneqq \img \coboundary\).
It is the orthogonal complement to the cycle space,
 \(\cycles^{\perp}=\cocycles\),
and we have
 \(\currents = \cycles \oplus \cocycles\)
reminiscent of the Helmholtz decomposition for vector fields in \(\reals^3\).

Also Kirchhoff's voltage law has an abstract analogue. 
The co-cycles \(y\in \cocycles\) satisfy
 \(\langle y, z\rangle = 0\)
for all \(z\in\cycles\).
In particular, they add up to zero when summing along any mesh of the graph.
Hence, co-cycles (like the voltages for electrical networks) correspond to the differences in potential functions defined on the edges.
The potentials are simply the elements \(u\in\potentials\).
The co-boundary operator \(\coboundary u\) then yields the corresponding ``voltages'' as their edge-wise differences.

In the next section, we provide a topological perspective to cycles of the graph that provides insight into the dimensions of the cycle and co-cycle spaces, 
and that admits the construction of nice bases for practical calculations.

\subsection{Topological approach to the cycle space}
\label{sec:topology}

In his pioneering work,  Hill~\citep{Hill1966,Hill1977} developed a cycle theory for  biological networks described by master equations.
It was later extended to a consistent network theory by Schnakenberg~\citep{Schnakenberg1976,Zia2007}.
In order to prepare the discussion of the gauge invariance of the fluctuation spectra in \Sec{gauge} we rephrase these concepts in the language of the algebraic graph theory introduced in the preceding subsection.

Two topologically special classes of graphs are circuits and trees:
a \emph{circuit} is a connected graph in which every vertex has exactly two neighbors.
Thus, for a circuit
 \( (\vertices, \oredges) \)
we have
 \( \abs{\vertices}=\abs{\oredges}\).
A \emph{tree} is a connected graph
 \( (\vertices, \tredges)\)
that does not contain any circuit as a sub-graph.
Consequently, every tree satisfies
 \( \abs{\tredges}=\abs{\vertices}-1\).

\begin{figure}
 \centering \includegraphics[scale=1]{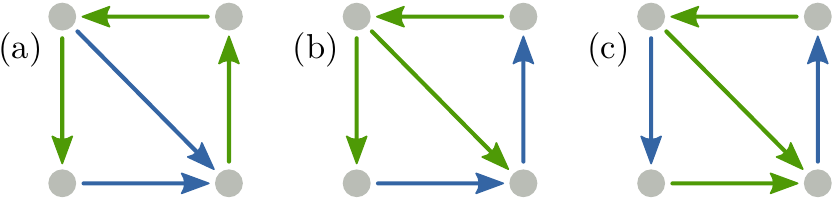}
 \caption{Three different spanning trees for the oriented graph given in \Fig{graph}(b).
  The edges \(\tredges\) of the different trees are marked green, while the chords \(\chords\) are depicted in blue.
  Up to symmetries, every other spanning tree of the graph looks like one of the depicted ones.
 }
 \label{fig:trees}
\end{figure}

Every connected graph
 \( (\vertices,\oredges) \)
contains a \emph{spanning tree}
 \( (\vertices,\tredges) \)
as a sub-graph.
This is a tree spanning all vertices and possibly less edges,
 \(\tredges\subseteq\oredges\).
In general, a graph contains many different spanning trees, as demonstrated in \Fig{trees}, where the edges that comprise the respective trees are indicated in green.
The edges
 \(\chords=\oredges\setminus\tredges\)
that are not part of the spanning tree are called \emph{chords} of the spanning tree.
These chords are colored in blue.

\begin{figure}
 \centering \includegraphics[scale=1]{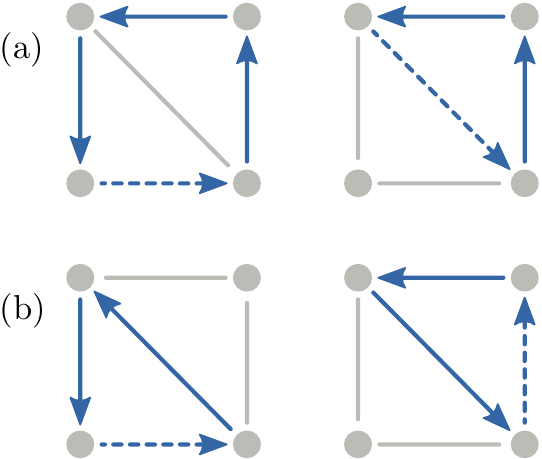}
 \caption{Fundamental cycles of the spanning trees in \Fig{trees}: every chord \(\eta\in\chords\) (dashed here, blue in \Fig{trees})
  generates a fundamental cycle, that is also marked in blue here.
  The gray edges 
  are not part of the fundamental cycles.
  Note that the spanning tree in \Fig{trees}(c) generates the exact same fundamental cycles as the one in \Fig{trees}(b), but with different chords.
  In contrast, the spanning tree in \Fig{trees}(a) shares only one fundamental cycle with the spanning trees (b) and (c).
 }
 \label{fig:cycles}
\end{figure}

Adding any chord \(\eta\in\chords\) to its spanning tree creates the sub-graph
 \( \left( \vertices,\tredges\cup\left\{ \eta \right\} \right)\).
This graph contains exactly one circuit.
Aligning all its edges in parallel to \(\eta\) and summing these, creates a so-called \emph{fundamental cycle} \(\zeta_\eta\in\cycles\).
The spanning trees in \Fig{trees} give rise to the fundamental cycles shown in \Fig{cycles}.
By construction, every fundamental cycle contains a different chord.
Thus, the fundamental cycles are linearly independent.
Moreover, in \App{fund-cocycles} we show that the fundamental cycles
 \(\left\{ \zeta_\eta\middle| \eta\in\chords \right\}\)
form a basis of the cycle space \(\cycles\), which we defined algebraically in \Sec{cycles-cocycles}.
Consequently, we have
 \(\abs{\chords}=\abs{\oredges}-\abs{\vertices}+1=\dim \cycles\),
irrespective of the choice of the spanning tree.
This number is a topological constant also known as \emph{cyclomatic number} or \emph{first Betti number}.
It immediately tells us, why trees and circuits are so important: trees have a trivial cycle space
 \(\cycles = \left\{ 0 \right\}\),
while circuits give rise to a one-dimensional cycle space.
In other words, cycles and trees are the topological building blocks of graphs.

Since the oriented edges
 \(\oredges=\chords\, \cup\, \tredges\)
are an ortho-normalized basis of \(\currents\), we have another orthogonal decomposition:
 \(\currents = \chordspace \oplus \treespace\).
The \emph{chord space} \( \chordspace \) and the \emph{tree space} \( \treespace\) depend on the choice of the spanning tree, while the definitions of \(\cycles\) and \(\cocycles\) are independent of the choice of a spanning tree.
Moreover, no single edge is a cycle nor a co-cycle.
Thus,
 \( \cycles \neq \chordspace\)
and
 \(\cocycles\neq\treespace\)
even though
 \( \dim \cycles =\abs{\chords}= \dim \chordspace\)
and
 \(\dim \cocycles = \abs{\tredges}= \dim \treespace \).

Note that the fundamental cycles are not ortho-normalized with respect to the standard scalar product:
 \( |\langle \zeta_{\eta_1}, \zeta_{\eta_2}\rangle| \)
counts the number of edges shared by the two fundamental cycles corresponding to the chords \(\eta_1\) and \(\eta_2\).
The sign of their scalar product indicates whether the shared edges are aligned parallel or anti-parallel.
Nonetheless, for a fundamental cycle \(\zeta\) and a chord \(\eta\) we have \[\langle \zeta, \eta\rangle = \zeta(\eta)= \begin{cases} 1\,, &
  \text{if } \zeta=\zeta_\eta\,,\\
  0\,, &
  \text{else}\,.\end{cases} \] 
Now we know that the fundamental cycles are a basis of the cycle space.
Thus, every cycle can be written as \( z = \sum_{\eta} z_{\eta}\, \zeta_{\eta}\).
Consequently, evaluating the cycle on a chord \(\eta'\) directly gives the coefficients of this linear combination: \(z(\eta')=\sum_{\eta}z_{\eta}\, \zeta_{\eta}(\eta') = z_{\eta'}\).
So, knowing the values of a cycle \(z\in\cycles\) on the chords \(\chords\subset\oredges\) alone is sufficient to reconstruct its values on all the edges \(\oredges\).
The most important special case is the steady-state probability current: \(J\in\cycles\).

\subsection{Summary}

We formulated the jump observables as a vector space \(\currents\), and showed that the multi-linearity entails a representation, \Eq{naive-multi-linear-cumulants}, of fluctuation spectra as superpositions of the spectra of the basis vectors of the vector space.
The vector space has two different natural decompositions into linear subspaces:
either universally as cycles and co-cycles \(\currents=\cycles\oplus\cocycles\) or with a given spanning tree as \(\currents=\reals^\chords\oplus\reals^\tredges\).
The former representation allowed us to formulate analogies (Kirchhoff's laws, Helmholtz decomposition) between electrodynamics and currents arising from Markov processes.
The latter allows the connection to the network theories of Hill and Schnakenberg \cite{Hill1966,Schnakenberg1976}.
In the following section we will use both representations to identify a gauge invariance of fluctuation spectra.
This gauge freedom substantially simplifies the calculation of fluctuation spectra introduced in Sec.~\ref{sec:currents}.

\section{Gauge Freedom of Fluctuation Spectra}
\label{sec:gauge}
Now everything is in place to state the fundamental result of the present work:
the fluctuation spectrum of an arbitrary observable \(\obs\in\currents\) only depends on its component in the cycle space {\(\cycles\).
On the one hand, this entails that observables share identical fluctuation spectra if they differ only by co-cyclic parts. 
On the other hand, this allows us to choose particularly easy representations.

The orthogonal decomposition of the edge space
 \(\currents=\cycles\oplus \cocycles\)
guarantees that every jump observable \(\obs\in\currents\) can be written as a unique sum \(\obs=z+y\) of a cycle \(z\in \cycles\) and a co-cycle \(y \in \cocycles\).
In view of the fact that the scaled cumulants are multi-linear, we can calculate the scaled cumulants of \(\obs\) from the scaled cumulants of \(z\) and \(y\).
In fact, the co-cyclic part can be neglected as stated in the following
\begin{Proposition*}
	\label{thm:main}
 The scaled cumulant-generating function \(\lambda_{\obs}(q)\) of a jump observable \(\obs=z+y\in\currents\)
 satisfies
  \(\lambda_{\obs}(q)=\lambda_z(q)\)
 where \(z\in \cycles\) and \(y\in\cocycles\) are the unique cycle and co-cycle parts of \(\obs\), respectively.
\end{Proposition*}
\begin{Proof}
 The matrix
  \(\mathbb{M}\coloneqq\W_{\obs}(q)-\lambda\mathbb{E}\)
 has entries
 \begin{align*}
  \mathbb{M}^i_j = \begin{cases} 	- \lambda -\sum_{k\neq i} w^i_k \,, &
   \text{if } i=j\,,\\
   w^i_j\, \exp\left({q\,\obs(e^i_j)}\right)\,, &
   \text{if } i\neq j\,.
  \end{cases}
 \end{align*}
 With the symmetric group \(\mathfrak{S}_N\), \ie{} the group of permutations of \(N\) symbols, we write the characteristic polynomial of \(\W_{\obs}(q)\) as
 \begin{align*}
  \chi_{\obs}(q,\lambda) = \det \mathbb{M} = \sum_{\sigma\in\mathfrak{S}_N} \sgn(\sigma)\, \mathbb{M}^1_{\sigma(1)} \mathbb{M}^2_{\sigma(2)}\dots \mathbb{M}^N_{\sigma(N)}\,.
 \end{align*}
 We will show that all contributions to
  \(\chi_{\obs}(q,\lambda)\)
 that have a dependence on \(\obs\) cannot distinguish between \(\obs\) and its cycle part \(z\), \ie{}
  \(\chi_{\obs}(q,\lambda)=\chi_z(q,\lambda)\).

 Every permutation
  \(\sigma\in\mathfrak{S}_N\)
 is a composition of several cyclic permutations of different lengths.
 There are  several cases of how a given permutation
  \(\sigma\in\mathfrak{S}_N\)
 contributes to the determinant:
 \begin{itemize}
 \item[(i)]
  There might be a state \(v_i\in\vertices\) such that \(w^{i}_{\sigma(i)}=0\).
  In that case the transition from \(v_i\) to \(v_{\sigma(i)}\) is not allowed, or equivalently, \(v_i\) and \(v_{\sigma(i)}\) are no neighbors.
  Then also
   \(\mathbb{M}^i_{\sigma(i)}=0\)
  and the entire summand in the determinant vanishes.
  In particular, the summand is independent of \(\obs\).
 \item[(ii)]
  Every fixed point of the permutation \(j=\sigma(j)\) contributes with \(\mathbb{M}^j_j\) which is independent of \(\obs\).
 \item[(iii)]
  For every neighboring transposition, \ie{} \(k=\sigma^2(k)\), we can use anti-symmetry,
   \(\obs(e^k_{\sigma(k)})+\obs(e^{\sigma(k)}_k)=0\),
  to conclude
   \(\mathbb{M}^k_{\sigma(k)}\mathbb{M}^{\sigma(k)}_k=w^k_{\sigma(k)}w^{\sigma(k)}_k\),
  which is independent of \(\obs\).
 \item[(iv)] Every remaining contribution must necessarily be a permutation along a properly oriented circuit of the graph.
  A properly oriented circuit has no boundary and consequently is a cycle.
  That means the summand in the determinant contains a product along a cycle \(\zeta\).
  In the exponent, this translates to summing the observable \(\obs\) along a cycle, \ie{} taking the scalar product
   \(\langle \obs, \zeta\rangle = \langle z, \zeta \rangle\),
   which is insensitive to the co-cyclic part of the observable.
 \end{itemize}
 Thus, we have shown that the characteristic polynomial only depends on the cyclic part:
  \(\chi_{\obs}(q,\lambda)=\chi_z(q,\lambda) \).
  Consequently, also its dominant root only depends on the cyclic part:
  \(\lambda_{\obs}(q)=\lambda_z(q)\).
\end{Proof}

Proposition~\ref{thm:main} establishes the ``gauge freedom'' referred to in the title.
We use this terminology for two reasons:
Firstly, it expresses the fact that observable results, \ie{} the scaled cumulants of time-averaged currents are the same for all jump observables in the subspace
 \(\obs+\cocycles\subset \currents\).
In particular, they agree with that of both \(\obs\) and its cycle part \(z\), \cf~\Fig{projection}.
Secondly, Polettini recently discussed various aspects of stochastic thermodynamics as a gauge theory, for both discrete~\citep{Polettini2012} and continuous~\citep{Polettini2013} situations.
In the discrete case, the gauge invariance he discusses is that of the jump observables $\sigma^i_j = \ln \frac{w^i_j}{w^j_i}$ and $\tilde{\sigma}^i_j=\ln \frac{\pi_iw^i_j}{\pi_jw^j_i}$ yielding the steady-state entropy production.
Their difference, \(\ln \frac{\pi_i}{\pi_j}\), is the gradient field $(\coboundary\ln{\pi})$.
Thus, it corresponds to a cocyclic gauge.
Here, we have established this invariance for arbitrary jump observables with any state-wise difference $y\in\cocycles$ serving as a gauge.

\begin{figure}
 \centering \includegraphics[scale=1.0]{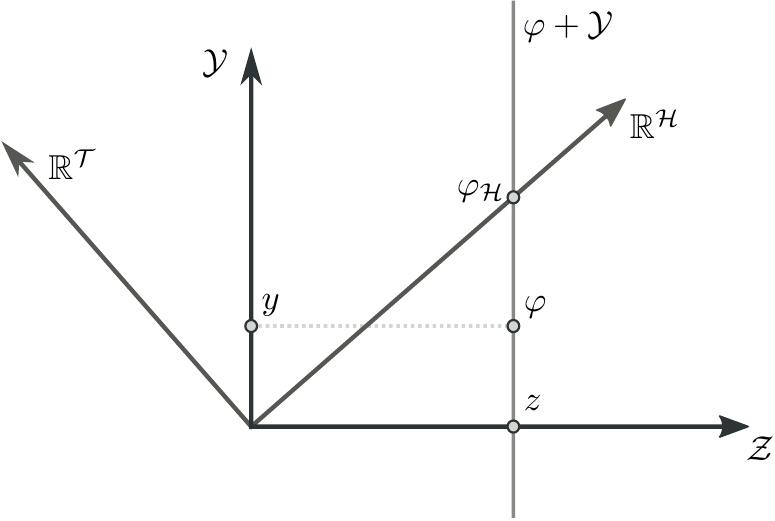}
 \caption{Geometrical interpretation of the chord representative: Projecting \(\obs\) in parallel to \(\cocycles\) onto the chord space \(\reals^\chords\) gives \(\obs_\chords\).
 }
 \label{fig:projection}
\end{figure}

The gauge freedom implies that one may choose any convenient representative in the class \(\obs+\cocycles\) to calculate the asymptotic statistics.
Often, the most convenient choice is neither \(\obs\) nor \(z\) but a representation that vanishes on some edges of the graph.
In Schnakenberg's network theory, similar considerations yield the Schnakenberg decomposition for the steady-state entropy production~\citep{Schnakenberg1976}.
The generalized notion for an arbitrary observable $\obs$ is its \emph{chord representation} 
\begin{align}
  \obs_{\chords}\coloneqq \sum_{\eta\in\chords} \langle \obs, \zeta_\eta\rangle\, \eta \equiv \sum_{\eta\in\chords} \circu{\obs}{\eta}\, \eta,
\end{align}
where
 \(\zeta_\eta\in \cycles\)
 is the fundamental cycle corresponding to the chord \(\eta\in\chords\) and \(\circu{\obs}{\eta}\coloneqq\langle\obs,\zeta_{\eta}\rangle\).
This special observable vanishes everywhere on the spanning tree.
The values it takes on the chords can be interpreted in the picture of vector fields:
the coefficient \(\circu{\obs}{\eta}\) is the ``circulation'' of \(\obs\) along a corresponding  fundamental cycle.
It is obtained by summing the values of $\obs$ for all edges that form the cycle $\zeta_\eta$ \cite{Altaner.etal2012}.
The circulation should not be confused with the value of the observable on the corresponding chord: \(\circu{\obs}{\eta} \neq \obs(\eta)\).
Geometrically speaking, the chord representation is the non-orthogonal (oblique) projection of \(\obs\) parallel to \(\cocycles\) onto \(\reals^\chords\), as depicted in \Fig{projection}.
Thus, \(\obs_\chords\) is the unique element in the intersection
 \( \chordspace \cap \left(\obs + \cocycles\right) \).
 These considerations are formally equivalent to using the non-orthogonal decomposition \(\currents=\reals^\chords\oplus\cocycles\), \cf{} Ref.~\citep{Polettini2015}.
As a consequence, the scaled cumulant-generating functions of \(\obs\) and \(\obs_\chords\) agree.
Moreover, \(\obs_\chords\) is the representative within \(\obs+\cocycles\) that is supported on a minimum number of edges.
This special gauge minimizes the effort to calculate the scaled cumulants.

Since the scaled cumulant-generating function satisfies
 \(\lambda_{\obs}(q)=\lambda_{\obs_{\chords}}(q)\),
\revi{we use the multi-linearity~(\ref{eq:naive-multi-linear-cumulants}) and the cycle representation in order to express the fluctuation spectrum of arbitrary observables by the fluctuation spectrum of chord cumulants.
More precisely, the $\nu$th cumulants $c(\obs^{(i_1)},\obs^{(i_2)}, \dots,\obs^{(i_\nu)})$ of a set of arbitrary observables are fully determined by their circulations $\mathring{\obs}^{(i_k)}_\eta$ and the $\nu$th cumulants of the chords, $c(\eta_1, \eta_2, \dots,\eta_\nu)$:}
\begin{subequations}
	\revi{
 \begin{align}
  c(\obs^{(i_1)}, \dots,\obs^{(i_\nu)}) &
  = \left( \prod_{k=1}^\nu \sum_{\eta_k\in\chords} \mathring{\obs}^{(i_k)}_{\eta_k}\right) c(\eta_1, \dots,\eta_\nu)
  \label{eq:gen-schnakenberg}\\
  \intertext{such that} c(\obs^{(\ell)}) &
  = \sum_{\eta\in\chords}\mathring{\obs}^{(\ell)}_{\eta}\, c(\eta)=\sum_{\eta\in\chords} \mathring{\obs}^{(\ell)}_{\eta}\, J(\eta)\,,\label{eq:schnakenberg}\\
 c(\obs^{(\ell)},\obs^{(m)}) &
 = \sum_{\eta_1\in\chords}\sum_{\eta_2\in\chords} \mathring{\obs}^{(\ell)}_{\eta_1} \mathring{\obs}^{(m)}_{\eta_2}\, c(\eta_1, \eta_2)\,.
 \end{align} }
 \label{eq:multi-linear-cumulants}%
\end{subequations}%

\revi{
\EQ{multi-linear-cumulants} separates the asymptotic fluctuation properties of the Markov process from the details of the (physical) observables $\obs^{(i_k)}$ under consideration.
This means that we do not need to apply the method based on the Implicit Function Theorem (Sec.~\ref{sec:analytics}) for each combination of observables individually.
Instead, we use it only to obtain the analytical expressions for the scaled cumulants \(c(\eta_1, \eta_2, \dots, \eta_{\nu})\) of the chord observables $\eta \in \chords\subset\currents$.
Remember that this can be done explicitly to any order, either by hand or using a computer algebra system (\cf{} Eq.~\eqref{eq:c1c2general} for the first two orders).
}

\revi{
\EQ{multi-linear-cumulants} states that the set of $\nu$th-order scaled cumulants \(c(\eta_1, \eta_2, \dots, \eta_{\nu})\) fully determines the fluctuation spectrum of \emph{arbitrary} observables at the order $\nu$.
The chord cumulants are a unique property of the Markov process, because the corresponding tilted generator depends on the transition matrix $\tmat$ only.
Cumulants of specific (physical) observables $\obs^{(i_k)}$ are obtained as multi-linear combinations. 
All the information required about an observable $\obs^{(i_k)}$ is its circulation $\mathring{\obs}^{(i_k)}_\eta$ on the fundamental cycles.
In our accompanying publication~\cite{AltanerWachtelVollmer2015}, we adopt this approach and present an algorithm to obtain the first (average currents) and second cumulants (correlation integrals) of arbitrary observables.}

\revi{The previous paragraph stresses} the importance of the circulations, or equivalently the chord representation, as the fundamental characteristic properties of jump observables.
\EQ{gen-schnakenberg} is the generalization of the Schnakenberg decomposition~\cite{Schnakenberg1976} for arbitrary observables and \emph{arbitrary orders of the fluctuation statistics}.
The choice of a spanning tree is part of the gauge freedom that may help to simplify the analysis of a jump observable \(\obs\in\currents\) even more:
in some cases a spanning tree might be chosen such that the chord representative \(\obs_{\chords}\) vanishes on a large number of chords.
Then the scaled cumulants of the fluctuating current \(\overline{\obs}_T\) involve a minimal number of joint cumulants of chord currents.
This can be seen as an optimal gauge for a given observable.

Finally, our result has practical significance for measuring the full asymptotic statistics of the entropy production $\sigma$.
For thermodynamically consistent models, the Hill--Schnakenberg conditions~\cite{Hill1966,Schnakenberg1976} relate the cycle affinities \(\circu{\sigma}{\eta}\) to thermodynamic drivings.
Then, the entropy production is most easily obtained via the chord representative $\sigma_\chords$, \cf{} our accompanying publication \cite{AltanerWachtelVollmer2015}.
In experiments one usually measures the statistics of individual transitions, and thus obtaines the fluxes $\pi_i w^i_j$ rather than the transition rates $w^i_j$.
In this situation the entropy production is accessible using $\tilde{\sigma}$ with entries $\ln{(\pi_iw^i_j/\pi_iw^i_j)}$ rather than $\sigma$ with entries $\ln{(w^i_j/w^i_j)}$.
Our result states that this distinction does not matter much:
the statistics from either choice $\sigma$, $\sigma_\chords$ or $\tilde{\sigma}$ converge to each other exponentially fast in time, which has been previously noted in simulations~\cite{Puglisi_etal2010}.

\section{Discussion and Conclusion}
\label{sec:discussion}

We have characterized the fluctuations of time-averaged currents in Markov processes in terms of the spectrum of the associated scaled cumulants.
They can be calculated by applying the Implicit Function Theorem to the characteristic equation \eq{characteristic} of the cumulant-generating function. 
Solely by taking derivatives and solving linear equations, one thus obtains analytic expressions \eq{c1c2general} for the scaled cumulants.
Due to the multi-linearity of cumulants and the vector-space structure of the observables one can express each scaled cumulant as a superposition, \Eq{naive-multi-linear-cumulants}, of the cumulants of a small set of observables whose currents form a basis of the vector space.
The vector space has two natural decompositions into linear subspaces:
a decomposition in terms of cycles and co-cycles \(\currents=\cycles\oplus\cocycles\) resembles the Helmholtz decomposition of vector fields in \(\reals^3\) and provides the connection to Kirchhoff's theory of electrical circuits;
a decomposition based on a spanning tree as \(\currents=\reals^\chords\oplus\reals^\tredges\) highlights the topological structure investigated by Hill and Schnakenberg.
With the former decomposition we identified a gauge invariance of the fluctuation spectra, Proposition 1.
This invariance, combined with the second decomposition, identifies the most effective representation, \Eq{multi-linear-cumulants}, of the scaled cumulants from the point of view of practical calculations.
Algebraically this representation is rooted in a combined non-orthogonal decomposition of the jump-observables as \(\currents=\reals^\chords\oplus\cocycles\), \cf{} \Fig{projection} and Ref.~\citep{Polettini2015}.

To the knowledge of the present authors, special cases to the general expressions in \Eqs{c1c2general} were first found in Ref.~\citep{Koza1999} for periodic systems.
A uni-variate version was also used in the Appendix to Ref.~\citep{BulnesCuetara.etal2011}.
The systematic theory for the multi-variate case and for statistics to arbitrary order was developed independently and concurrently in Refs.~\citep{Wachtel2013,Bruderer2014}. 

For the first cumulant, \ie{} the expectation value, the representation in \Eq{schnakenberg} is a generalization of the \emph{Schnakenberg decomposition} for the dissipation rate \citep{Schnakenberg1976} to general jump observables.
\EQ{gen-schnakenberg} generalizes this decomposition for all orders of the fluctuation statistics and all jump observables -- it stresses the importance of the circulations, or equivalently the chord representation, as the fundamental characteristic properties of jump observables.
For the dissipation rate, these circulations are the cycle affinities that play the central role in the theory developed by \citet{Hill1977} and \citet{Schnakenberg1976}.
For general observables, their importance in the cyclic decomposition was addressed in a previous publication~\citep{Altaner.etal2012}.
The results presented here emphasize the role of the cycle topology of a network of states.
Preserving the circulations of observables is also important in the context of coarse-graining of stochastic models~\citep{Altaner.Vollmer2012}.

A first instance of the connection of Schnakenberg's network theory to statistics is the work by Andrieux and Gaspard~\citep{Andrieux2007}:
They proved a Gallavotti--Cohen type symmetry for the currents along the chords.
This was later slightly extended to other bases in the cycle space by~\citet{Faggionato2011}.
The present work establishes that not only the Gallavotti--Cohen symmetry, but that all the large-deviation properties depend on the cycles alone.

There are also other approaches to access the fluctuation spectra analytically~\citep{2005EPL-Flindt.etal,2009JoSP-Baiesi.etal}.
These methods rely on analytical knowledge of the steady-state distribution \(\vec{\pi}\).
In principle this unique distribution exists as long as the Markov process is ergodic.
Nonetheless, it is very difficult to find analytic expressions for \(\vec{\pi}\) if the transition matrix \(\W\) depends on parameters in an analytical way and this dependence should be carried over into the steady state.
On the contrary, using our method to calculate first the steady-state probability currents \(J\), and then to calculate the steady-state probability distribution \(\pi\) based on \(J\) and \(\W\) proved to be a lot faster in our experience, \cf{} Ref.~\citep{AltanerWachtelVollmer2015}. 

\revi{In addition to analytical methods, there are advanced numerical methods to calculate the large deviation properties of the entropy flow~\citep{Imparato.Peliti2007}, general current-like observables~\citep{Giardina.etal2006} and other classes of observables~\citep{Chetrite.Touchette2013}.
  Numerical approaches, however, intrinsically have the disadvantage that they must be re-evaluated for every change of parameters and every change of observable.
  This makes systematic parameter scans and (possibly in addition) the comparison of several observables computationally very costly. 
  The approach presented here overcomes this limitation and is especially efficient for low orders of the fluctuation spectrum, \ie{} the statistics that can be measured reliably in experiments, \cf{}~the accompanying work~\cite{AltanerWachtelVollmer2015} for a more thorough discussion.
}

In conclusion, we established an efficient algorithm to characterize the fluctuations of time-averaged currents in finite Markov processes.
We explicitly stress the following points:
\begin{itemize}
  \item The asymptotic fluctuation spectrum of every physical current can be calculated analytically without the need to solve non-linear equations, \cf~\Sec{analytics}.
  \item Due to the multi-linearity of the scaled cumulants, one can first calculate the fluctuation spectrum of a basis set of fundamental observables.
    The scaled cumulants of other observables then amount to appropriate multi-linear combinations, \Eqs{naive-multi-linear-cumulants}.
  \item Every jump observable can be written as a unique sum of a cycle and a co-cycle, 
    and asymptotic fluctuations (as characterized by Large Deviation Theory) of arbitrary currents admit a co-cyclic gauge, Proposition~\ref{thm:main}:
    the co-cyclic part of the observables does not contribute to fluctuations.
  \item This gauge yields a representation of the fluctuation spectrum in terms of scaled cumulants of fundamental cycle contributions, \Eqs{multi-linear-cumulants}.
    An informed choice of the cycles leads to an efficient algorithm for determining 
    all scaled cumulants.
\end{itemize}

The formalism introduced here admits a systematic and efficient calculation of all scaled cumulants for all observables of arbitrary finite, dynamically-reversible Markov processes. 
In Stochastic Thermodynamics, such models are commonly used as models of molecular motors, \ie{} macromolecules which facilitate bio-chemical processes in living cells \cite{Seifert2008,Seifert2012}.
Hence, our results facilitate the comparison of experiments and the predictions of Stochastic Thermodynamics, as well as pinpointing differences of the predictions of different models.
In an accompanying paper \citep{AltanerWachtelVollmer2015} we demonstrate the power of our method for the study of models parametrized by physical driving forces.
In particular, we report on the molecular energy transduction between different non-equilibrium reservoirs and the nonequilibrium response theory of the molecular motor kinesin. 

\subsection*{Acknowledgements}

The authors thank Matteo Polettini, Fabian Telschow, and Hugo Touchette for insightful discussions.
Florian Angeletti provided useful comments on the manuscript.
AW is grateful for a Ludwig Prandtl internship awarded by FOKOS~e.\,V, and to M. Esposito for his support.
JV acknowledges a research grant of the  ``Center for Earth System Research and Sustainability'' while the final version of this manuscript was drafted.

\begin{figure}
 \centering \includegraphics[scale=1]{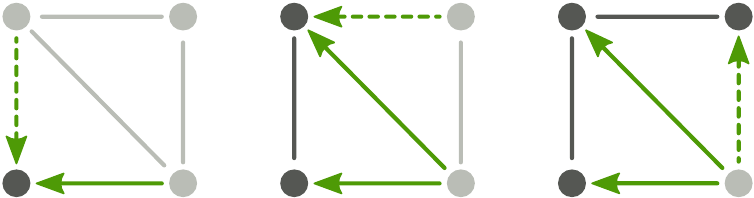}
 \caption{Fundamental co-cycles of the spanning tree in \Fig{trees}a:
  removing an edge \(\tau\in\tredges\) (dashed) from the spanning tree, decomposes the graph into two disconnected components marked in two different shades of gray.
  The edges marked in green form the corresponding fundamental co-cycle: they are the set of edges pointing from one component to the other.
 }
 \label{fig:cocycles}
\end{figure}
\appendix

\section{Dimension of the Cycle Space and Fundamental Co-cycles}
\label{sec:fund-cocycles}

The space of jump observables has an orthogonal decomposition as \(\currents = \cycles \oplus \cocycles\), \cf{} \Sec{cycles-cocycles}.
We will now elaborate on the dimensions of these orthogonal subspaces.
As pointed out in \Sec{topology}, every chord \(\eta\in\chords\) of a spanning tree \(\tredges\) defines a fundamental cycle \(\zeta_\eta\).
We already established that the fundamental cycles have vanishing boundary and thus are elements of the cycle space, \(\zeta_\eta\in\cycles\).
Moreover, they are linearly independent.
Here we give a short argument, why they are a basis of the cycle space.
A more detailed account on the methods of algebraic topology applied to graph theory can be found in~\citep{Knauer2011}.

Our argument starts from an additional set of independent vectors that lie in the co-cycle space~\(\cocycles\):
Removing an edge \(\tau\in\tredges\) from the spanning tree of a graph separates the latter into two components, the smallest possible component being a single vertex without any edge.
The set \(\chords\cup\{\tau\}\) contains a minimal subset that separates these two components and it is therefore called \emph{cut}.
Reorienting the edges of this cut to be parallel to \(\tau\) and summing these edges, results in the \emph{fundamental co-cycle} corresponding to the removed edge \(\tau\), \cf{} the examples in \Fig{cocycles}.

It is easy to check that every fundamental co-cycle indeed is a co-cycle: it is the co-boundary of a scalar field \(u\in\potentials\) taking the value \(1\) on the one disconnected component and the value \(0\) on the other.
Moreover, each fundamental co-cycle contains a different edge of the spanning tree and thus these co-cycles are linearly independent.

Obviously, the total number of fundamental cycles and co-cycles equals the total number of edges in the graph.
Consequently, the fundamental cycles and co-cycles together are a maximal set of linearly independent vectors and thus span the space \(\currents\) of jump observables.
This implies that the fundamental cycles are a basis for the cycle space while the fundamental co-cycles are a basis of the co-cycle space.
This results in \(\dim \cycles = \abs{\chords} = \abs{\oredges}-\abs{\vertices} + 1\) and \(\dim \cocycles = \abs{\tredges} = \abs{\vertices} -1\).

Note that this reasoning is independent of the choice of a spanning tree.
The spanning tree, however, defines the cycles and co-cycles that are regarded fundamental.

\bibliography{current_cumulants}

\end{document}